\documentstyle[pra,aps]{revtex}
\begin{document}

\begin{center}
{\huge Phenomenological--operator approach to introduce damping effects on
radiation field states}$^{\dagger }${\huge \ }

N. G. de Almeida, P. B. Ramos, R. M. Serra, and M. H. Y. Moussa*

{\it Departamento de F\'{i}sica, Universidade Federal de S\~{a}o Carlos, CP.
676, S\~{a}o Carlos 13565-905, SP, Brazil}

\bigskip

{\bf Abstract}

\smallskip
\end{center}

\begin{quote}
In this work we propose an approach to deal with radiation field states
which incorporates damping effects at zero temperature. By using some well
known results on dissipation of a cavity field state, obtained by standard
ab-initio methods, it was possible to infer through a phenomenological way
the explicit form for the evolution of the state vector for the whole
system: the cavity-field plus reservoir. This proposal turns out to be of
extreme convenience to account for the influence of the reservoir over the
cavity field. To illustrate the universal applicability of our approach we
consider the attenuation effects on cavity-field states engineering. The
main concern of the present phenomenological approach consists in furnishing
a straightforward technique to estimate the fidelity resulting from
processes in cavity QED phenomena. A proposal to maximize the fidelity of
the process is presented.

\smallskip
\end{quote}

PACS number(s): 32.80.-t, 42.50.Dv

{\it \ * E-mail: miled@power.ufscar.br}

$\dagger $ {\it to appear in Journal of Optics B: Quantum and Semiclassical
Optics}

\section{Introduction}

After more than half a century since their proposition by Einstein,
Podolsky, and Rosen (EPR) \cite{epr}, entangled states have become the
cornerstone of a set of striking proposals in theoretical physics. The
advent of Bell's theorem \cite{bell}, which has permitted to test
empirically the phenomenon of nonlocality, and more than a decade of
experiments confirming such an astonishing character of quantum mechanics
gave to entanglement the necessary credibility to motivate such proposals.
Ranging from quantum communication \cite{zoller} and cryptography \cite{chau}
to teleportation\cite{bennett,zeilinger} and quantum computation \cite{shor}%
, the nonlocal character of entangled states has let its original realm of
quantum metaphysics to inaugurate the possibility of a new technology. In
fact, inspired by these theoretical implications, the experimental setups,
formerly designed to investigate fundamental physics such as quantum
coherence and nonlocality, have recently been enhanced for the realization
of teleportation \cite{zeilinger} and to demonstrate quantum-logic
operations \cite{wineland}.

However, the above-mentioned processes on quantum mechanics come up against
a crucial problem intrinsic to quantum nature: the {\it decoherence} coming
from the inevitable coupling between quantum systems and the surrounding
environment. Such a decoherence process, transforming a pure state $\left|
\psi \right\rangle $ in a statistical mixture $\rho $, constitutes the main
difficulty preventing the realization of the above mentioned quantum
processes. Since the transformation $\left| \psi \right\rangle $ $%
\longrightarrow $ $\rho $ becomes faster as the excitation of the initial
quantum state increases, the realization of massive computation, for
example, turns out to be prohibitive and even the realization of a single
logic gate operation \cite{wineland} leads to a degraded output. In this
connection, the calculation of the fidelity of a given process becomes a
crucial task when proposing its experimental realization. However, this task
turns out to be as difficult to achieve as the implementation of the
protocol for accomplishing a given quantum process. As an example, we cite
the process of engineering a cavity-field state, which requires a series of
steps to be pursued \cite{papers}, since in most of these schemes the
cavity-field is built up photon by photon.

To overcome the difficulty in estimating the fidelity in cavity QED
processes, in this paper we propose a phenomenological way to handle with
dissipation of a cavity-field state under the influence of a reservoir at
absolute zero temperature. Our strategy is to incorporate, in an concise
algebraic approach, the main results obtained by standard {\it ab-initio}
techniques on dissipation of a cavity field. In short, we include the
exponential decay law accounting for the damping of the cavity-field
excitation explicitly in the evolution of the whole system comprehending the
cavity field plus reservoir. Once we have calculated the evolved whole state
vector, it is immediate to obtain the total density matrix, and by tracing
out the reservoir variables we finally get the desired reduced density
matrix for the cavity field, which is the same as the one obtained by
standard methods. This approach resembles the Monte Carlo Wave Function
method \cite{mcwf} in the sense that we work directly with the wave
function, providing an efficient computational tool to account for
dissipation in quantum optics.

It is worth mentioning that our result for the evolved state vector of the
radiation field plus reservoir can be applied to calculate dissipation
effects in any physical process based on cavity QED. Here, as an
application, we consider damping effects in engineering a quantum field
state and present an alternative method to maximize the fidelity of the
engineered state. The process of generation of an arbitrary radiation field
state, as will be here analyzed, is severely damaged due to the reservoir
attenuation.

This paper is organized as follows. In section II we describe the essential
features of the model here considered, a monochromatic cavity-field trapped
in a lossy cavity. In section III we present an algebraic way to introduce
damping explicitly on the evolution of the whole state vector for\ the
cavity field plus reservoir. In section IV we apply the technique developed
in section III to analyze the fidelity of an engineered cavity-field state.
Section V is devoted to illustrate a technique to optimize the fidelity of
the engineered field state and finally, in section VI we present our
conclusions.

\section{Model}

We are concerned with the standard model of a cavity field described by
creation and annihilation central oscillator amplitude operators $a^{\dagger
}$, $a$, coupled to a reservoir consisting of a set of oscillators described
by amplitude operators $b_{k}^{\dagger }$, $b_{k}$. The reservoir simulates
the cavity damping mechanism: photon losses either at the mirrors or by
leaking out of the cavity. The cavity field is coupled to the reservoir
through the usual rotating wave approximation, so that the Hamiltonian for
the system is written as 
\begin{equation}
H=\hbar \omega a^{\dagger }a+\sum_{k}\hbar \omega _{k}b_{k}^{\dagger
}b_{k}+\sum_{k}\hbar \left( g_{k}ab_{k}^{\dagger }+g_{k}^{\ast }a^{\dagger
}b_{k}\right) ,  \label{hamiltoniano}
\end{equation}
where $g_{k}$ are coupling parameters. The Heisenberg equations for both
amplitude operators $a$ and $b$ are given by 
\begin{mathletters}
\begin{eqnarray}
\stackrel{.}{a}(t) &=&-i\omega a-i\sum_{k}g_{k}b_{k},  \label{1a} \\
\stackrel{.}{b_{k}}(t) &=&-i\omega _{k}b_{k}-ig_{k}^{\ast }a.  \label{1b}
\end{eqnarray}
Considering the Wigner-Weisskopf approximation, in which the frequencies of
the reservoir oscillators densely cover a range in the neighborhood of the
field frequency, we obtain the solutions of the equations corresponding to
the system (\ref{1a},\ref{1b}). The solution for the cavity field reads 
\end{mathletters}
\begin{equation}
a(t)=\mu (t)a(0)+\sum_{k}\vartheta _{k}(t)b_{k}(0).  \label{heisenberg}
\end{equation}
Introducing the damping factor $\gamma $ and the typical small frequency
shift $\Delta \omega $, the function $\mu (t)$ is explicitly given by 
\[
\mu (t)=\exp \left\{ -\left[ \frac{\gamma }{2}+i\left( \Delta \omega +\omega
\right) \right] t\right\} . 
\]
For the purpose at hands it is not necessary to specify the functions $%
\vartheta _{k}(t)$.

A convenient way to calculate the reduced density matrix for the cavity
field consists in using the standard characteristic function technique \cite
{livros}. For this intent, we must first calculate the characteristic
function $\chi $, then the Glauber-Sudarshan $P$-representation, and finally
the reduced density operator. The characteristic function can be written in
terms of the operators $a$, $a^{\dagger }$ in the normal order form as 
\begin{mathletters}
\begin{eqnarray}
\chi (\eta ,\eta ^{\ast },t) &=&\text{Tr}\left\{ \rho (t)\exp \left[ \eta
a^{\dagger }(0)\right] \exp \left[ -\eta ^{\ast }a(0)\right] \right\}
\label{funcao qui a} \\
&=&\text{Tr}\left\{ \rho (0)\exp \left[ \eta a^{\dagger }(t)\right] \exp %
\left[ -\eta ^{\ast }a(t)\right] \right\} .  \label{funcao qui b}
\end{eqnarray}
Eq.(\ref{funcao qui a}) refers to the Schr\"{o}dinger picture and the Eq. (%
\ref{funcao qui b}) to the Heisenberg picture. Since the time evolution of
operators $a,a^{\dagger }$ are known from Eq.(\ref{heisenberg}) together
with the initial state of the whole system, Eq.(\ref{funcao qui b}) turns
out to be more convenient to calculate the characteristic function.
Considering that the cavity field is initially in an arbitrary superposition
state $\sum_{n=0}^{N}C_{n}\left| n\right\rangle $, where $\left|
n\right\rangle $ is a $n$-photon Fock state, and the reservoir (assumed at
absolute zero temperature) is in the vacuum state, the total density matrix
at $t=0$ is given by 
\end{mathletters}
\begin{equation}
\rho (0)=\sum_{n,m=0}^{N}C_{n}C_{m}^{\ast }\left| n\right\rangle
\left\langle m\right| \otimes \left| \left\{ 0\right\} \right\rangle
\left\langle \left\{ 0\right\} \right| ,  \label{ro inicial}
\end{equation}
where we have denoted $\prod_{k}\left| 0_{k}\right\rangle \equiv \left|
\left\{ 0\right\} \right\rangle $.

The Glauber-Sudarshan $P$-representation of the density matrix can be
written as a two-dimensional Fourier transform of the characteristic
function $\chi $, 
\begin{equation}
P(\alpha ,\alpha ^{\ast },t)=\int \frac{d^{2}\eta }{\pi ^{2}}\chi (\eta
,\eta ^{\ast },t)\exp \left( -\eta \alpha ^{\ast }+\eta ^{\ast }\alpha
\right) ,  \label{representacao p}
\end{equation}
while the reduced density operator for the cavity field is given by, in the
general form, 
\begin{equation}
\rho (t)=\int d^{2}\alpha P(\alpha ,\alpha ^{\ast },t)\left| \alpha
\right\rangle \left\langle \alpha \right| .
\end{equation}

Considering all these definitions it is possible to obtain, after a little
algebra, the explicit form for the reduced density operator of the cavity
field 
\begin{eqnarray}
\rho (t) &=&\sum_{n,m=0}^{N}\sum_{j=0}^{m}\sum_{l=0}^{j}C_{n}C_{m}^{\ast
}\left( -1\right) ^{l}\frac{\left| \mu (t)\right| ^{2j}\mu (t)^{n-m}}{%
l!\left( m-j\right) !}  \nonumber \\
&&\sqrt{\frac{n!m!}{\left( j+n-l-m\right) !\left( j-l\right) !}}\left|
j+n-l-m\right\rangle \left\langle j-l\right| .  \label{ro final}
\end{eqnarray}

\section{Phenomenological approach}

The aim of this section is to include explicitly the energy dissipation of
the cavity field in the time evolution of the state vector for the whole
system. In short, the exponential decay law accounting for the damping of
the cavity-field excitation will be explicitly included into the evolution
of the state vector for the whole system. Such a state vector must lead,
naturally, to the result derived from formal techniques for the reduced
density matrix Eq.(\ref{ro final}). As usual, the reduced density matrix for
the cavity field is obtained by tracing over the reservoir variables, i.e., 
\begin{equation}
\rho _{S}(t)=\text{Tr}_{R}\left[ \rho _{S+R}(t)\right] ,  \label{traco1}
\end{equation}
where $\rho _{S+R}(t)$ is given, for pure states at time $t$, by 
\begin{equation}
\rho _{S+R}(t)=\left| \Psi _{S+R}\left( t\right) \right\rangle \left\langle
\Psi _{S+R}\left( t\right) \right| ,  \label{traco2}
\end{equation}
and the subscripts $S$ and $R$ refer to the system and reservoir,
respectively. What is thus necessary to be known is precisely the evolution
of the whole state vector $\left| \Psi _{S+R}\left( t\right) \right\rangle $
. For this intent, we begin by considering first some examples which will
guide us to the form of the evolution of the generalized $n-$photons
cavity-field state in the presence of the reservoir. For the simplest case,
a single photon in the cavity, the evolution of the whole state vector can
be phenomenologically inferred as 
\begin{equation}
\left| 1\right\rangle \left| R\right\rangle \longrightarrow \exp \left( -%
\frac{\gamma }{2}t\right) \left| 1\right\rangle \left| R\right\rangle
+\left| 0\right\rangle \sum_{k}\alpha _{k}b_{k}^{\dagger }\left|
R\right\rangle ,  \label{primeiro}
\end{equation}
where each $\alpha _{k}$ represents the probability amplitude that a photon
have been absorbed by the $k$-th mode of the reservoir, and $\gamma $ is the
damping rate for the one-photon state in the cavity. Under the normalization
condition on state (\ref{primeiro}), we arrive at the constraint 
\begin{equation}
\sum_{k}\left| \alpha _{k}\right| ^{2}=1-\exp \left( -\gamma t\right) .
\label{constraint}
\end{equation}

For the next case of a two-photon state, we infer the evolution of the whole
system by noting that the interaction between the cavity mode and the
reservoir is always characterized by one-photon exchange. In this way, the
final form of the evolution of the whole state vector can be guessed as 
\begin{eqnarray}
\left| 2\right\rangle \left| R\right\rangle &\longrightarrow &\exp \left( -2%
\frac{\gamma }{2}t\right) \left| 2\right\rangle \left| R\right\rangle + 
\nonumber \\
&&\sqrt{2}\exp \left( -\frac{\gamma }{2}t\right) \left| 1\right\rangle
\sum_{k}\alpha _{k}b_{k}^{\dagger }\left| R\right\rangle +  \nonumber \\
&&\left| 0\right\rangle \sum_{kk^{\prime }}\alpha _{k}\alpha _{k^{\prime
}}b_{k}^{\dagger }b_{k^{\prime }}^{\dagger }\left| R\right\rangle .
\label{segunda}
\end{eqnarray}
The first term on the r.h.s of Eq.(\ref{segunda}) is associated with the
probability that the cavity field remains in its original two-photon state.
It is worth noting the difference between the relaxation rates for one- and
two-photon field states: the damping rate for a two-photon state is
increased by a factor $2$ compared to that of the one-photon state. In the
second term on the r.h.s of Eq.(\ref{segunda}), we have included the
probability that the one-photon state has originated from the decay of the
two-photon state and also the probability that the one-photon state remains
as such. The last term accounts for the probability that all photons have
been absorbed by the reservoir. We stress that the imposition of the
normalization condition on state Eq.(\ref{segunda}) gives the same
constraint Eq.(\ref{constraint}).

We note that states Eqs.(\ref{primeiro}) and (\ref{segunda}), as it should
be, recover the same result as given by Eq.(\ref{ro final}).\ To verify this
assertion, it is necessary to build up the density operator for the system
plus reservoir and after that to get rid of the reservoir variables.

Finally it is possible, by induction, to generalize the above results to
construct the evolution of an arbitrary $n$-photon state. This state vector
for the whole system includes the ingredients discussed before and can be
read as 
\begin{eqnarray}
\left| n\right\rangle \left| R\right\rangle &\longrightarrow &\exp \left( -n%
\frac{\gamma }{2}t\right) \left| n\right\rangle \left| R\right\rangle 
\nonumber \\
&&+\sqrt{\frac{n!}{1!(n-1)!}}\exp \left[ -\left( n-1\right) \frac{\gamma }{2}%
t\right] \left| n-1\right\rangle \sum_{k_{1}}\alpha
_{k_{1}}b_{k_{1}}^{\dagger }\left| R\right\rangle  \nonumber \\
&&+\sqrt{\frac{n!}{2!(n-2)!}}\exp \left[ -\left( n-2\right) \frac{\gamma }{2}%
t\right] \left| n-2\right\rangle \sum_{k_{1}k_{2}}\alpha _{k_{1}}\alpha
_{k_{2}}b_{k_{1}}^{\dagger }b_{k_{2}}^{\dagger }\left| R\right\rangle 
\nonumber \\
&&+...+\left| 0\right\rangle \sum_{k_{1}...k_{n}}\alpha _{k_{1}}...\alpha
_{k_{n}}b_{k_{1}}^{\dagger }...b_{k_{n}}^{\dagger }\left| R\right\rangle ,
\label{geral}
\end{eqnarray}
where each term appearing in Eq.(\ref{geral}) includes the probability that
a given state has originated from the initial $n$-photon state and also the
probability that it will remain the same. Again, the desired reduced density
matrix for the cavity field can be obtained from the time-dependent state
vector Eq$.$(\ref{geral}) by using Eqs.(\ref{traco1}) and (\ref{traco2}).

\section{ Noise effects on engineering quantum-field states}

To illustrate the applicability of the method developed in the previous
section, we restrict our analysis to the damping effects in the engineering
process of a cavity-field state. To do that, we choose Vogel et al.'s scheme
to engineer an arbitrary cavity-field state\cite{vogel}, which is based on
successive resonant interactions of $M$ two-level atoms with an initially
empty cavity. The experimental setup for this scheme is depicted in Fig.1.
The Rydberg atoms are laser excited before entering into the Ramsey zone $R$%
, placed in the way of the atoms to the cavity $C$, preparing each atom in a
particular superposition of excited $\left| e\right\rangle $ and ground $%
\left| g\right\rangle $ states, as is required to properly build up the
cavity field, photon by photon. Each two-level atom , after interacting with
a monochromatic field state in the cavity $C$, leaves its photon in the
cavity, being necessarily detected in its ground state by detector chambers $%
D$ ($D_{e}$ and $D_{g}$ for ionizing the states $\left| e\right\rangle $ and 
$\left| g\right\rangle $, respectively).

To exemplify an application of the present phenomenological approach, we
consider the first step of this process. Initially, the cavity field $C$ is
in the vacuum state while the atom is in an arbitrary (unnormalized)
superposition $\left| e\right\rangle $ $+i\epsilon _{1}$ $\left|
g\right\rangle $, where $\epsilon _{1}$ is a complex parameter controlled by
the Ramsey zone $R$. Once we are considering the reservoir at absolute zero
temperature, it will always be in the vacuum state $\left| R\right\rangle
=\left| \left\{ 0\right\} \right\rangle $.

To simplify the calculations, we assume a stable excited atomic state and
just take into account the errors introduced by the cavity dissipation
mechanism. This is a good approximation, since on average eight out of ten
atoms are able to travel the distance of the whole setup without decaying.
In fact, since a Rydberg-atom excited state has a lifetime $\left( 1/\gamma
_a\right) $ of the order of $10^{-2}$ s\cite{brune}, the probability of
staying in this state is about $0.8$ for an experiment duration of about $%
2\times 10^{-3}$s. For high-Q superconducting cavities, the lifetime $\left(
1/\gamma \right) $ is also of the order of $10^{-2}$s \cite{brune}. Finally,
we neglect the field dissipation during the time that the atoms interact
with the cavity field.

After the resonant atom-field interaction, when the atom has left the cavity
and measured in the ground state, the state of the whole system can be
written as 
\begin{equation}
\left| \Psi ^{(1)}\right\rangle =\sum_{n=0}^{1}\widehat{Q}_{n}^{(1)}\left|
n\right\rangle \left| R\right\rangle ,  \label{1}
\end{equation}
where we have described the damping effects on the cavity field (given by
Eq.(\ref{primeiro})) through the operators $\widehat{Q}_{n}^{(1)}$ appearing
above, which are explicitly given by 
\begin{mathletters}
\begin{eqnarray}
\widehat{Q}_{0}^{(1)} &=&\varphi _{0}^{(1)}+\varphi _{1}^{(1)}\sum_{k}\alpha
_{k}^{\left( 1\right) }b_{k}^{\dagger },  \label{2a} \\
\widehat{Q}_{1}^{(1)} &=&\exp \left( -\frac{\gamma }{2}t^{\prime }\right)
\varphi _{1}^{(1)},  \label{2b}
\end{eqnarray}
where $t^{\prime }$ is the time in which the relaxation process occurs,
i.e., the time interval in which the first atom left the cavity and is
detected. At this point it is useful to define the parameters $%
C_{n}^{(k)}=\cos (g\tau _{k}\sqrt{n+1})$ and $S_{n}^{(k)}=\sin (g\tau _{k}%
\sqrt{n+1})$, where $\tau _{k}$ is the interaction time of the $k$-th atom
with the cavity field and $g$ is the atom-field coupling constant, assumed
to be the same for all the atoms. The constants $\varphi _{i}^{(1)}$, $i=1,2$
in Eqs.(\ref{2a},\ref{2b}) are the same as those in Ref.\cite{vogel}, i.e., 
\end{mathletters}
\begin{eqnarray}
\varphi _{1}^{(1)} &=&S_{0}^{(1)}, \\
\varphi _{0}^{(1)} &=&-\epsilon _{1}.
\end{eqnarray}

The next step consists in the analysis of the results after the passage of
the second atom through the cavity and the subsequent relaxation of the
radiation field then constructed. After the second atom (prepared in the
superposition $\left| e\right\rangle $ $+i\epsilon _{2}$ $\left|
g\right\rangle $) has left the cavity and again been measured in the ground
state, the two-photon cavity field-reservoir state becomes 
\begin{equation}
\left| \Psi ^{(2)}\right\rangle =\sum_{n=0}^{2}\widehat{Q}_{n}^{(2)}\left|
n\right\rangle \left| R\right\rangle ,  \label{fi}
\end{equation}
where the operators $\widehat{Q}_{n}^{(2)}$ turns out to be more complicated
then those in Eq.(\ref{1}) due to entanglement between the states of each
atom and the cavity field-reservoir states. These operators can be written
as 
\begin{mathletters}
\begin{eqnarray}
\widehat{Q}_{0}^{(2)} &=&\widehat{T}_{0}^{(2)}+\widehat{T}%
_{1}^{(2)}\sum_{k}\alpha _{k}^{\left( 2\right) }b_{k}^{\dagger }+\widehat{T}%
_{2}^{(2)}\sum_{k_{1}k_{2}}\alpha _{k_{1}}^{\left( 2\right) }\alpha
_{k_{2}}^{\left( 2\right) }b_{k_{1}}^{\dagger }b_{k_{2}}^{\dagger },
\label{4a} \\
\widehat{Q}_{1}^{(2)} &=&\widehat{T}_{1}^{(2)}\exp \left( -\frac{\gamma }{2}%
t\right) +\sqrt{2}\widehat{T}_{2}^{(2)}\exp \left( -\frac{\gamma }{2}%
t^{\prime }\right) \sum_{k}\alpha _{k}^{\left( 2\right) }b_{k}^{\dagger },
\label{4b} \\
\widehat{Q}_{2}^{(2)} &=&\widehat{T}_{2}^{(2)}\exp \left( -2\frac{\gamma }{2}%
t\right) ,  \label{4c}
\end{eqnarray}
where the operators $\widehat{T}_{n}^{(2)}$appearing above are given by 
\end{mathletters}
\begin{mathletters}
\begin{eqnarray}
\widehat{T}_{0}^{(2)} &=&-\epsilon _{2}\widehat{Q}_{0}^{(1)},  \label{5a} \\
\widehat{T}_{1}^{(2)} &=&S_{0}^{(2)}\widehat{Q}_{0}^{(1)}-\epsilon
_{2}C_{0}^{(2)}\widehat{Q}_{1}^{(1)},  \label{5b} \\
\widehat{T}_{2}^{(2)} &=&S_{1}^{(2)}\widehat{Q}_{1}^{(1)}.  \label{5c}
\end{eqnarray}
Here $t$ is the time starting after the second atom left the cavity. For
simplicity, we have assumed the time interval for the detections of the
first and second atoms as equal, and that the second atom enters into the
cavity immediately after the first atom being detected. \bigskip Although
here we are just concerned with the desired $\left| \psi _{d}\right\rangle
=\sum_{n=0}^{2}d_{n}\left| n\right\rangle $ cavity field state, the method
described above can be generalized for generating an arbitrary cavity field
state ( $\left| \psi _{d}\right\rangle =\sum_{n=0}^{M}d_{n}\left|
n\right\rangle $) in a straightforward (if rather tedious) manner.
Naturally, all the informations we need are contained in $\left| \Psi
^{(2)}\right\rangle $ according to Eq. (\ref{fi}). For instance, the reduced
density operator can be obtained readily by using Eq. (\ref{fi}) and
definitions Eqs. (\ref{traco1}) and (\ref{traco2}) as

\end{mathletters}
\begin{eqnarray}
\widehat{\rho }_{S}(t) &=&{\cal N}\left[ a\left| 2\right\rangle \left\langle
2\right| +b\left| 1\right\rangle \left\langle 1\right| +c\left|
0\right\rangle \left\langle 0\right| +\right.  \label{rosistema} \\
&&\left. +\left( d\left| 2\right\rangle \left\langle 1\right| +f\left|
2\right\rangle \left\langle 0\right| +g\left| 1\right\rangle \left\langle
0\right| +h.c.\right) \right] ,  \nonumber
\end{eqnarray}
where the coefficients appearing above are given in appendix $A$, and ${\cal %
N}=1\left/ \left( a+b+c\right) \right. $ is the normalization constant.

From this reduced density operator, Eq. (\ref{rosistema}), which includes
the inevitable coupling between the system and the reservoir, it is possible
to calculate, in principle, any physical quantity (observable) of interest.

{\it Ideal case. }When damping effects are neglected, i.e., letting the
damping constant $\gamma =0$, Eq. (\ref{rosistema}) recovers the ideal case,
and $\widehat{\rho }_{S}$ can be written in terms of a pure state $\left|
\Psi _{S}\right\rangle \left\langle \Psi _{S}\right| $, where $\left| \Psi
_{S}\right\rangle =\sum_{n=0}^{2}\psi _{n}^{(2)}\left| n\right\rangle $ and
the coefficients $\psi _{n}^{(2)}$ are the same as those in Ref. \cite{vogel}%
, as can be checked from Eqs.(\ref{2a}) to (\ref{5c}).

\section{Optimizing the fidelity of an engineered cavity-field state}

At this point this formulation is completely general, and can be applied to
generate any arbitrary cavity field state. However, we stress that there is
a crucial difference between the procedure described here and the one in
Ref. \cite{vogel}. In fact, in the treatment proposed by Vogel et al., it is
necessary to solve a polynomial equations for $\epsilon _{1\text{ }}$and $%
\epsilon _{2}$, which arise from the consistence of the recurrence equations
relating the amplitudes $\varphi _{k}^{(n)}$. On the other hand, in the
treatment described here, in virtue of the reservoir, the quantities $%
\widehat{Q}_{n}^{(2)}$ in Eq. (\ref{fi}) are now operators, instead of
simply being parameters. In this connection, there is no immediate way to
extract useful information from the recurrence relation associated between
the operators involved. Thus, a first question arising is, how to turn round
this problem? We propose the following way to solve this problem: consider a
specific desired cavity-field state to be engineered and calculate the
fidelity of this ideal state (free from the reservoir effects) with respect
to the state $\left| \Psi ^{(2)}\right\rangle $ in Eq. (\ref{fi}). By doing
this procedure, the only unknown variables are $\epsilon _{1}$ and $\epsilon
_{2}$, since all the others can be fixed. The trick here is to find
numerically the values for $\epsilon _{1}$and $\epsilon _{2}$ which maximize
the resulting expression for the fidelity. This process can be even more
optimized by an adequate choice of the interaction parameters $g\tau _{k}$.

Another important ingredient in this discussion is concerned with the
probability for successfully achieve the engineering process. In fact, the
success of the engineering process depends on detecting all the required
atoms in their ground states. The expression for the probability to detect
each atom in the ground state can be obtained in an usual way. The
distinguished feature here is given by the presence of the reservoir. To do
that, we have to consider the internal atomic states as well as the
constructed radiation-field states being taken normalized. We thus have to
normalize the states given in Eqs. (\ref{1},\ref{fi}) by following the recipe%
$\left| \Psi ^{(k)}\right\rangle \rightarrow {\cal N}_{k}\left| \Psi
^{(k)}\right\rangle $, where the normalization constants read

\begin{mathletters}
\begin{eqnarray}
{\cal N}_{1} &=&\left[ \frac{1}{1+\left| \epsilon _{1}\right| ^{2}}%
\left\langle R\right| \sum_{j=0}^{1}\left( \widehat{Q}_{j}^{(1)}\right)
^{\dagger }\widehat{Q}_{j}^{(1)}\left| R\right\rangle \right] ^{-\frac{1}{2}%
},  \label{norma} \\
{\cal N}_{2} &=&\left[ \frac{1}{1+\left| \epsilon _{2}\right| ^{2}}\frac{%
\left\langle R\right| \sum_{j}\left( \widehat{Q}_{j}^{(2)}\right) ^{\dagger }%
\widehat{Q}_{j}^{(2)}\left| R\right\rangle }{\left\langle R\right|
\sum_{j}\left( \widehat{Q}_{j}^{(1)}\right) ^{\dagger }\widehat{Q}%
_{j}^{(1)}\left| R\right\rangle }\right] ^{-\frac{1}{2}}.  \label{normb}
\end{eqnarray}

The total probability${\cal \ }$for successfully engineering the desired
state reads ${\cal P=}\prod_{k=1}^{2}{\cal P}_{k}$, where ${\cal P}_{k}=$ $%
1/\left( {\cal N}_{k}\right) ^{2}$ accounts for the probability to detect
the $k$-th atom in the ground state $\left| g\right\rangle $.

To exemplify the whole procedure, we restrict our analysis to the generation
of the following radiation-field truncated phase-state,

\end{mathletters}
\[
\left| \Psi _{d}\right\rangle =\frac{1}{\sqrt{3}}\left( \left|
0\right\rangle +\left| 1\right\rangle +\left| 2\right\rangle \right) . 
\]
The fidelity${\cal \ }$of the density matrix $\rho _{s}(t)$ relative to $%
\left| \Psi _{d}\right\rangle $ is given by

\begin{equation}
{\cal F}=\left\langle \Psi _d\right| \rho _s(t)\left| \Psi _d\right\rangle ,
\label{fidel}
\end{equation}
where $\rho _s(t)$ is the normalized density matrix. The resulting
expression is a little bit involved and must be numerically maximized
regarding the variables $\epsilon _1$ and $\epsilon _2.$ To optimize the
engineering process (\ref{fidel}), we consider different interaction
parameters $g\tau _k$ to each atom, which depend on experimental
capabilities. In the present realistic scheme for engineering a cavity-field
state two important features arise: the fidelity ${\cal F}$ of the
engineered state and the probability ${\cal P}$ for successfully achieve the
process. Fig.2(a,b,c) show the histogram of the probability ${\cal P}$, the
fidelity ${\cal F}$ , and the rate ${\cal R\equiv PF}$, which serves as a
cost-benefit estimate to choose the best parameters, in terms of the
parameters $g\tau _1$ and $g\tau _2$. For a definite value of the
interaction parameters $g\tau _k$, it is calculated the correspondent values 
$\epsilon _1$ and $\epsilon _2$ which maximize the fidelity. From this
figure we note that a larger probability does not necessarily imply a better
fidelity. This is an important point since the convenient probability ${\cal %
P}$ and the fidelity ${\cal F}$ to be chosen depend on the engineer
necessities. In Table I some values of $g\tau _k$ and the solutions for $%
\epsilon _1$ and $\epsilon _2$ as well as the correspondent quantities $%
{\cal P}$, ${\cal F}$ , and ${\cal R}$ are exhibited. Fig.3(a-f) show both
the histogram of the elements of the reduced density matrix and the
associated Wigner functions. Fig.3(a,b) show the ideal case ($\gamma =0$);
Fig.3(c,d) show the realistic engineered state by using the parameters $%
\epsilon _1$ and $\epsilon _2$ obtained from Vogel et. al scheme taking into
account the reservoir; and Fig.3(e,f) show the engineered state obtained by
our maximization procedure. As can be viewed in Fig.3(e,f), this procedure
improve the quantum nature of the engineered state, which is revealed by the
negative portions of the Wigner function. To plot the figures 2 and 3 we
used the realistic experimental parameters $\gamma =10^2s^{-1}$ \cite{brune}
for the decay rate and the estimated time interval $t=t^{\prime }=10/\gamma $
\cite{paulo} for the first and second atoms, respectively, to reache the
detection chambers after leaving the cavity.

\section{Conclusion}

In this paper we have proposed an alternative way to treat damping effects
at zero temperature in cavity QED processes. Our phenomenological approach
consists in incorporate the main results obtained by standard {\it ab-initio}
techniques on dissipation of a cavity field directly in the evolution of the
state vector of the whole system: cavity field plus reservoir. In summary,
we have included the exponential decay law accounting for the damping of the
cavity-field excitation explicitly in the evolution of the whole state
vector of the system.

By considering a standard model of a given mode of the radiation field
coupled to a collection of $N$-modes representing the reservoir, it was
possible to infer the time evolution for the whole state vector in a
straightforward manner. We stress that this result is rather general, and in
principle can be applied to whichever quantum process as, for instance,
quantum communication, logic operations, teleportation, and cavity-field
state engineering, among others.

The phenomenological-operator approach to dissipation in cavity quantum
electrodynamics here presented considerably simplifies the introduction of
the inevitable errors due to the environmental degrees of freedom when
describing processes involving atom-field interactions. The development of
this technique became possible \ due to the previous formal work in quantum
dissipation, from which we have recovered the main results concerning energy
loss of a trapped radiation field. So, such convenient approach precludes
the necessity of performing the usually extensive ab initio calculations as
the standard master equation, the characteristic functions, or even the path
integral formalism. It is worth noting that the technique here developed at
absolute zero can be extended for a thermal reservoir.

As an application of the present technique, we have considered the process
of engineering a cavity-field state in a lossy cavity. In order to estimate
how far the generated state deviates from the idealized, due to attenuation,
we have analyzed the fidelity of the engineering process. Also, we have
proposed an alternative way for engineering a cavity field state in the
presence of a reservoir, which consists in maximizing the fidelity of the
desired (ideal) state relative to the (damped) engineered state. A specific
example has been given which takes into account realistic values for the
parameters involved. To conclude, we should stress that damping effects
seriously restrict such an engineering process. However, these noise effects
can be minimized by the scheme proposed in this paper.

\bigskip

{\bf Acknowledgments}

We wish to thank J.R.G. de Mendon\c{c}a for useful comments on this
manuscript and the support from Fapesp and CNPq, Brazil.

\bigskip

{\LARGE Appendix A}

In this appendix we show explicitly the coefficients appearing in Eq. (\ref
{rosistema}),

\[
a=\left( S_{0}^{(1)}S_{1}^{(2)}\right) ^{2}e^{-\gamma (t^{\prime }+2t)}, 
\]

\begin{eqnarray*}
b &=&2\left( S_{0}^{(1)}S_{1}^{\left( 2\right) }\right) ^{2}e^{-\gamma
(t^{\prime }+t)}\left( 1-e^{-\gamma t}\right) + \\
&&-\left| \epsilon _{2}\right| ^{2}\left( S_{0}^{\left( 1\right)
}C_{0}^{\left( 2\right) }\right) ^{2}e^{-\gamma (t^{\prime }+t)}+ \\
&&+\left( S_{0}^{\left( 1\right) }S_{0}^{\left( 2\right) }\right)
^{2}e^{-\gamma t^{\prime }}\left( 1-e^{-\gamma t}\right) + \\
&&-\left| \epsilon _{1}\right| ^{2}\left( S_{0}^{\left( 2\right) }\right)
^{2}e^{-\gamma t}+ \\
&&+\left( \epsilon _{1}\epsilon _{2}^{*}+\epsilon _{1}^{*}\epsilon
_{2}\right) S_{0}^{\left( 1\right) }C_{0}^{\left( 2\right) }S_{0}^{\left(
2\right) }e^{-1/2\gamma (t^{\prime }+2t)},
\end{eqnarray*}

\begin{eqnarray*}
c &=&-\left| \epsilon _{2}\right| ^{2}\left( S_{0}^{\left( 1\right) }\right)
^{2}\left( 1-e^{-\gamma t^{\prime }}\right) +\left| \epsilon _{1}\right|
^{2}\left| \epsilon _{2}\right| ^{2}+ \\
&&+\left( S_{0}^{\left( 1\right) }S_{1}^{\left( 2\right) }\right)
^{2}e^{-\gamma t^{\prime }}(1-e^{-\gamma t})^{2}+ \\
&&-\left| \epsilon _{2}\right| ^{2}\left( S_{0}^{\left( 1\right)
}C_{0}^{\left( 2\right) }\right) ^{2}e^{-\gamma t^{\prime }}\left(
1-e^{-\gamma t}\right) + \\
&&+\left( S_{0}^{(1)}S_{0}^{(2)}\right) ^{2}\left( 1-e^{-\gamma t^{\prime
}}\right) \left( 1-e^{-\gamma t}\right) + \\
&&-\left| \epsilon _{1}\right| ^{2}\left( S_{0}^{(2)}\right) ^{2}\left(
1-e^{-\gamma t}\right) + \\
&&+\left( \epsilon _{1}\epsilon _{2}^{*}+\epsilon _{1}^{*}\epsilon
_{2}\right) S_{0}^{(1)}S_{0}^{(2)}C_{0}^{(2)}e^{-1/2\gamma t^{\prime
}}\left( 1-e^{-\gamma t}\right) ,
\end{eqnarray*}

\[
d=-\epsilon _{2}^{*}\left( S_{0}^{(1)}\right)
^{2}S_{1}^{(2)}C_{0}^{(2)}e^{-\gamma (t^{\prime }+3/2t)}-\epsilon
_{1}^{*}S_{0}^{(1)}S_{1}^{(2)}S_{0}^{(2)}e^{-\gamma /2(t^{\prime }+3t)}, 
\]

\[
f=\epsilon _{1}^{*}\epsilon _{2}^{*}S_{0}^{(1)}S_{1}^{(2)}e^{-\gamma
/2(t^{\prime }+2t)}, 
\]

\begin{eqnarray*}
g &=&-\epsilon _{2}^{\ast }\left( S_{0}^{(1)}\right)
^{2}S_{0}^{(2)}e^{-1/2\gamma t}\left( 1-e^{-\gamma t^{\prime }}\right) + \\
&&-\epsilon _{1}^{\ast }\left| \epsilon _{2}\right| ^{2}S_{0}^{\left(
1\right) }C_{0}^{\left( 2\right) }e^{-\gamma /2(t^{\prime }+t)}+ \\
&&-\epsilon _{2}^{\ast }\left| \epsilon _{1}\right| ^{2}S_{0}^{\left(
2\right) }e^{-1/2\gamma t}+ \\
&&-\sqrt{2}\epsilon _{2}^{\ast }\left( S_{0}^{(1)}\right)
^{2}S_{1}^{(2)}C_{0}^{(2)}e^{-\gamma /2(2t^{\prime }+t)}\left( 1-e^{-\gamma
t}\right) + \\
&&-\sqrt{2}\epsilon _{1}^{\ast }S_{0}^{(1)}S_{0}^{(2)}S_{1}^{(2)}e^{-\gamma
t^{\prime }}\left( 1-e^{-\gamma t}\right) .
\end{eqnarray*}

{\bf Figure Caption}

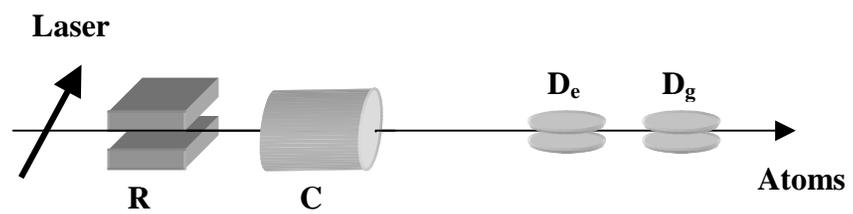
FIG. 1. Sketch of the experimental setup for engineering a cavity-field
state.

FIG. 2(a,b,c). Histogram of the probability ${\cal P}$, fidelity ${\cal F}$,
and rate ${\cal R}$, respectively, in the $g\tau _{1}\times $ $g\tau _{2}$
plane.

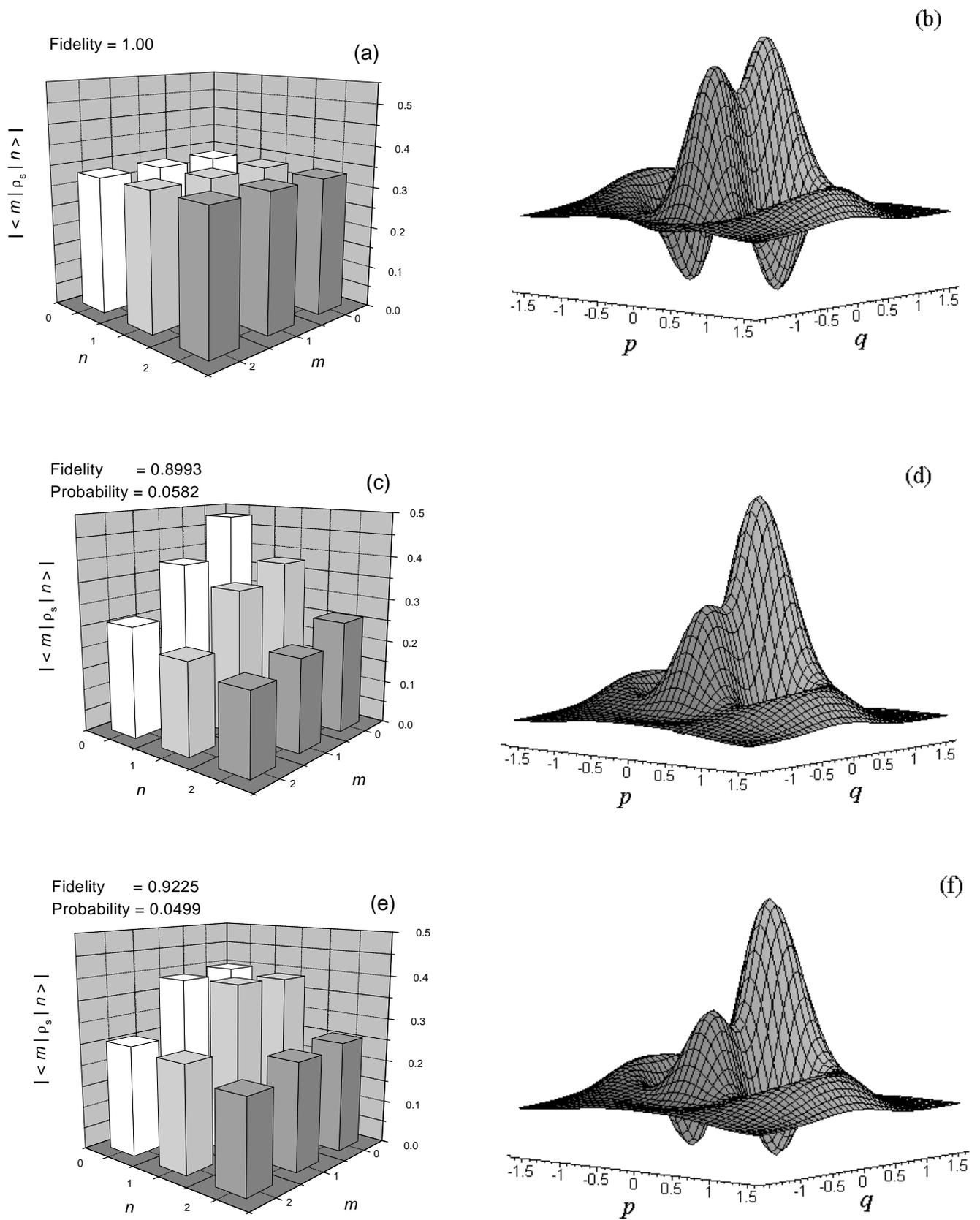
FIG. 3. Elements of the reduced density matrix and the correspondents Wigner
distribution functions: (a,b) for the ideal state ; (c,d) for the realistic
engineered state built with parameters $\epsilon _{1}$ and $\epsilon _{2}$
obtained from Vogel et al. scheme including the reservoir; and (e,f) for the
optimized engineered state by the maximization procedure. The Wigner $%
W\left( \alpha ,\alpha ^{\ast }\right) $ distribution functions are plotted
as a function of $q=%
%TCIMACRO{\func{Re}}%
%BeginExpansion
\mathop{\rm Re}%
%EndExpansion
(\alpha )$ and $p=%
%TCIMACRO{\func{Im}}%
%BeginExpansion
\mathop{\rm Im}%
%EndExpansion
(\alpha )$.

\bigskip

\bigskip

\noindent {\bf Table}

Table 1. Some values of the interaction times $g\tau _{1}$ and $g\tau _{2}$,
Ramsey zones parameters $\epsilon _{1}$ and $\epsilon _{2}$ (adjusted in
order to maximize the fidelity), fidelity ${\cal F}$, probability ${\cal P}$%
, and rate ${\cal R}$.

\begin{tabular}{ccccccc}
\hline\hline
$g\tau _{1}$ & $g\tau _{2}$ & $\quad \quad ~\quad \epsilon _{1}\quad ~~~~$ & 
$\quad \quad \quad ~\quad \epsilon _{2}\quad ~~~~$ & $\quad \quad ~{\cal F}%
~~~~$ & $\quad {\cal P}~~~~~$ & $~{\cal R}~~$ \\ \hline
\end{tabular}

\begin{tabular}{ccccccc}
$0.6$ & $3.0$ & $2.7693$ & $-0.1583$ & $0.9253$ & $0.0697$ & $0.0645$ \\ 
$1.3$ & $1.2$ & $-0.8508+i0.2874$ & $-0.8838-i0.1600$ & $0.8789$ & $0.8831$
& $0.7761$ \\ 
$1.4$ & $2.8$ & $1.2349+i0.8632$ & $-0.3583+i0.2427$ & $0.9087$ & $0.3637$ & 
$0.3305$ \\ \hline\hline
\end{tabular}


\begin{references}
\bibitem{epr}  A. Einstein, B. Podolsky, N. Rosen, Phys. Rev.{\bf \ 47}, 777
(1935).

\bibitem{bell}  J. S. Bell, Physics (N.Y.) {\bf 1}, 195 (1964).

\bibitem{zoller}  J. I. Cirac, P. Zoller, H. J. Kimble, and H. Mabuchi,
Phys. Rev. Lett. {\bf 78}, 3221 (1997); T. Pellizzari, Phys. Rev. Lett. {\bf %
79}, 5242 (1997).

\bibitem{chau}  H.-K. Lo and H. Chau, P. W. Shor, Phys. Rev. Lett. {\bf 78},
3410 (1997). D. Mayers, quant-ph/9603015.

\bibitem{bennett}  C. H. Bennett, G. Brassard, C. Cr\'{e}peau, R. Jozsa, A.
Peres, and W. Wootters, Phys. Rev. Lett. {\bf 70}, 1895 (1993); M. H. Y.
Moussa, Phys. Rev. A {\bf 54}, 4661 (1996); {\it ibid.} {\bf 55}, R3287
(1997); D. Bouwmeester, J.-W. Pan, K. Mattle, M. Eibl, H. Weinfurter, and A.
Zeilinger, Nature {\bf 390}, 575 (1997); D. Boschi, S. Branca, F. De
Martini, L. Hardy, and S. Popescu, Phys. Rev. Lett. {\bf 80}, 1121 (1998);
C. J. Villas-B\^{o}as, N. G. de Almeida, and M. H. Y. Moussa, Phys. Rev. A 
{\bf 60}, 2759 (1999).

\bibitem{zeilinger}  A. Furusawa, J. L. Sorensen, S. L. Braunstein, C. A.
Fuchs, H. J. Kimble, and E. S. Polzik, Science {\bf 282, }706 (1998), and
references therein.

\bibitem{shor}  P. W. Shor, in {\it Proceedings of the 35th Annual Symposium
on the Foundations of calculater Science}, edited by S. Goldwasser (IEEE
calculater Society Press, Los Alamitos, CA, (1994), p. 124.

\bibitem{wineland}  I. L. Chuang, L. M. K. Vandersypen, X. Zhou, D. W.
Leung, and S. Lloyd, {\it Nature}, {\bf 393}, 143 (1998), and references
therein.

\bibitem{papers}  J. Krause, M. O. Scully, H. Walther, Phys. Rev. A {\bf 36 }%
R4547 (1987); M. S. Brune, S. Haroche, J. M. Raimond, L. Davidovich, and N.
Zagury, Phys. Rev. A {\bf 45}, 5193 (1992); A. S. Parkins, P. Marte, P.
Zoller, and H. J. Kimble, Phys. Rev. Lett. {\bf 71} 3095 (1993); J. Janszky,
P. Domokos, S. Szab\'{o}, and P. Adam, Phys. Rev. A {\bf 51}, 4191 (1995);
S. Bose, K. Jacobs, and P. L. Knight Phys. Rev. A {\bf 56}, 4175 (1997); K.
M. Gheri, C. Saavedra, P. T\"{o}rm\"{a}, J. I. Cirac, and P. Zoller, Phys.
Rev. A {\bf 58} R2627 (1998); M. H. Y. Moussa, B. Baseia, Phys. Lett. A {\bf %
238} 223 (1998); Shi-Biao Zheng, Opt. Commun.{\bf \ 154} 290 (1998); A.
Vidiella-Barranco and J. A. Roversi, Phys. Rev. A {\bf 58}, 3349 (1998); D.
T. Pegg, L. S. Phillips, and S. M. Barnett, Phys. Rev. Lett {\bf 81}, 1604
(1998); M. Dakna, J. Clausen, L. Kn\"{o}ll, and D. G. Welsch, Phys. Rev. A 
{\bf 59} 1658 (1999); D. Branning, W. P. Grice, R. Erdmann, and I. A.
Walmsley, Phys. Rev. Lett. {\bf 83}, 955 (1999).

\bibitem{mcwf}  J. Dalibard, Y. Castin, and K. M\o lmer, Phys. Rev. Lett 
{\bf 68}, 580 (1992); R. Dum, P. Zoller, and H. Ritsch, Phys. Rev. A {\bf 45}%
, 4879 (1992).

\bibitem{vogel}  K. Vogel, V. M. \ Akulin, and W. P. Schleich, Phys. Rev.
Lett {\bf 71}, 1816 (1993).

\bibitem{livros}  See for example, M. O. Scully and M. S. Zubairy,{\it \ in
Quantum Optics}, Cambridge University Press (1997); L. Mandel and E. Wolf, 
{\it in Optical Coherence and Quantum Optics}, Cambridge University Press
(1995).

\bibitem{brune}  M. Weindinger, B. T. H. Varcoe, R. Heerlein, and H.
Walther, Phys. Rev. Lett. {\bf 82}, 3795 (1999); M. Brune, E. Hagley, J.
Dreyer, X. Ma\^{i}tre, A. Maali, C. Wunderlich, J. M. Raimond, and S.
Haroche, Phys. Rev. Lett. {\bf 77}, 4887 (1996).

\bibitem{paulo}  P. Nussenzveig, {\it private communication.}
\end{references}
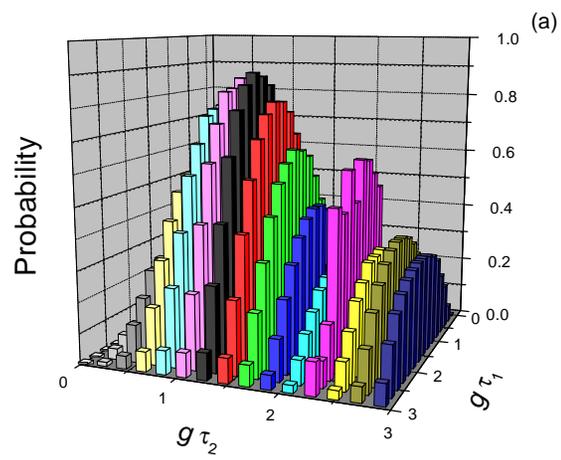
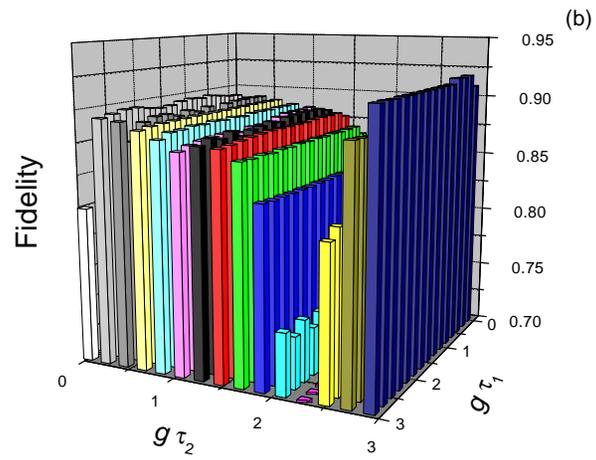
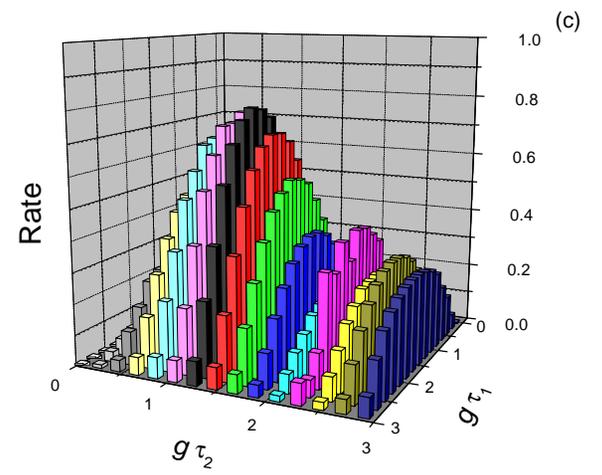
\end{document}